\documentstyle[12pt]{article}

\def\hybrid{\topmargin -20pt    \oddsidemargin 0pt
        \headheight 0pt \headsep 0pt
        \textwidth 6.25in       
        \textheight 9.5in       
        \marginparwidth .875in
        \parskip 5pt plus 1pt   \jot = 1.5ex}

\hybrid

\def\baselinestretch{1.2}

\catcode`\@=11

\def\marginnote#1{}
\newcount\hour
\newcount\minute
\newtoks\amorpm
\hour=\time\divide\hour by60
\minute=\time{\multiply\hour by60 \global\advance\minute by-\hour}
\edef\standardtime{{\ifnum\hour<12 \global\amorpm={am}%
        \else\global\amorpm={pm}\advance\hour by-12 \fi
        \ifnum\hour=0 \hour=12 \fi
        \number\hour:\ifnum\minute<10 0\fi\number\minute\the\amorpm}}
\edef\militarytime{\number\hour:\ifnum\minute<10 0\fi\number\minute}

\def\draftlabel#1{{\@bsphack\if@filesw {\let\thepage\relax
   \xdef\@gtempa{\write\@auxout{\string
      \newlabel{#1}{{\@currentlabel}{\thepage}}}}}\@gtempa
   \if@nobreak \ifvmode\nobreak\fi\fi\fi\@esphack}
        \gdef\@eqnlabel{#1}}
\def\@eqnlabel{}
\def\@vacuum{}
\def\draftmarginnote#1{\marginpar{\raggedright\scriptsize\tt#1}}

\def\draft{\oddsidemargin -.5truein
        \def\@oddfoot{\sl preliminary draft \hfil
        \rm\thepage\hfil\sl\today\quad\militarytime}
        \let\@evenfoot\@oddfoot \overfullrule 3pt
        \let\label=\draftlabel
        \let\marginnote=\draftmarginnote
   \def\@eqnnum{(\theequation)\rlap{\kern\marginparsep\tt\@eqnlabel}%
\global\let\@eqnlabel\@vacuum}  }

\def\preprint{\twocolumn\sloppy\flushbottom\parindent 2em
        \leftmargini 2em\leftmarginv .5em\leftmarginvi .5em
        \oddsidemargin -.5in    \evensidemargin -.5in
        \columnsep .4in \footheight 0pt
        \textwidth 10.in        \topmargin  -.4in
        \headheight 12pt \topskip .4in
        \textheight 6.9in \footskip 0pt
        \def\@oddhead{\thepage\hfil\addtocounter{page}{1}\thepage}
        \let\@evenhead\@oddhead \def\@oddfoot{} \def\@evenfoot{} }

\def\numberbysection{\@addtoreset{equation}{section}
        \def\theequation{\thesection.\arabic{equation}}}

\def\underline#1{\relax\ifmmode\@@underline#1\else
        $\@@underline{\hbox{#1}}$\relax\fi}

\def\titlepage{\@restonecolfalse\if@twocolumn\@restonecoltrue\onecolumn
     \else \newpage \fi \thispagestyle{empty}\c@page\z@
        \def\thefootnote{\fnsymbol{footnote}} }

\def\endtitlepage{\if@restonecol\twocolumn \else \newpage \fi
        \def\thefootnote{\arabic{footnote}}
        \setcounter{footnote}{0}}  

\catcode`@=12
\relax

\def\figcap{\section*{Figure Captions\markboth
        {FIGURECAPTIONS}{FIGURECAPTIONS}}\list
        {Figure \arabic{enumi}:\hfill}{\settowidth\labelwidth{Figure
999:}
        \leftmargin\labelwidth
        \advance\leftmargin\labelsep\usecounter{enumi}}}
 \relax
\def\tablecap{\section*{Table Captions\markboth
        {TABLECAPTIONS}{TABLECAPTIONS}}\list
        {Table \arabic{enumi}:\hfill}{\settowidth\labelwidth{Table
999:}
        \leftmargin\labelwidth
        \advance\leftmargin\labelsep\usecounter{enumi}}}
 \relax
\def\reflist{\section*{References\markboth
        {REFLIST}{REFLIST}}\list
        {[\arabic{enumi}]\hfill}{\settowidth\labelwidth{[999]}
        \leftmargin\labelwidth
        \advance\leftmargin\labelsep\usecounter{enumi}}}
 \relax
\makeatletter
\newcounter{pubctr}
\def\publist{\@ifnextchar[{\@publist}{\@@publist}}
\def\@publist[#1]{\list
        {[\arabic{pubctr}]\hfill}{\settowidth\labelwidth{[999]}
        \leftmargin\labelwidth
        \advance\leftmargin\labelsep
        \@nmbrlisttrue\def\@listctr{pubctr}
        \setcounter{pubctr}{#1}\addtocounter{pubctr}{-1}}}
\def\@@publist{\list
        {[\arabic{pubctr}]\hfill}{\settowidth\labelwidth{[999]}
        \leftmargin\labelwidth
        \advance\leftmargin\labelsep
        \@nmbrlisttrue\def\@listctr{pubctr}}}
 \relax
\makeatother
\newskip\humongous \humongous=0pt plus 1000pt minus 1000pt

\newif\ifdtup

\relax

\def\be{\begin{equation}}
\def\ee{\end{equation}}
\def\ba{\begin{eqnarray}}
\def\ea{\end{eqnarray}}

\def\no{\noindent}

\def\IR{\relax{\rm I\kern-.18em R}}

\def\IR{\relax{\rm I\kern-.18em R}}
\def\inv{^{\raise.15ex\hbox{${\scriptscriptstyle -}$}\kern-.05em 1}}

\begin{document}

\renewcommand{\theequation}{\arabic{equation}}

\newcommand{\beq}{\begin{equation}}
\newcommand{\eeq}[1]{\label{#1}\end{equation}}
\newcommand{\ber}{\begin{eqnarray}}
\newcommand{\eer}[1]{\label{#1}\end{eqnarray}}
\newcommand{\eqn}[1]{(\ref{#1})}
\begin{titlepage}
\begin{center}

\hfill hep--th/0312274\\
\vskip -.1 cm
\hfill December 2003\\

\vskip .7in

{\large \bf Ricci flows and infinite dimensional algebras}\footnote{Contribution
to the proceedings of the {\em 36th International Symposium Ahrenshoop on the Theory 
of Elementary Particles}, Berlin, August 26--30, 2003; to be published in 
Fortschritte der Physik.}

\vskip 0.6in

{\bf Ioannis Bakas}
\vskip 0.2in
{\em Department of Physics, University of Patras \\
GR-26500 Patras, Greece\\
\footnotesize{\tt bakas@ajax.physics.upatras.gr}}\\

\end{center}

\vskip .8in

\centerline{\bf Abstract}

\no
The renormalization group equations of two-dimensional sigma models describe  
geometric deformations of their target space when the world-sheet length    
changes scale from the ultra-violet to the infra-red. These equations, which are  
also known in the mathematics literature as Ricci flows, are analyzed for the  
particular case of two-dimensional target spaces, where they are found to admit a 
systematic description as Toda system. Their zero curvature formulation is made
possible with the aid of a novel infinite dimensional Lie algebra, which has 
anti-symmetric Cartan kernel and exhibits exponential growth. The general solution
is obtained in closed form using B\"acklund transformations, and special 
examples include the sausage model and the decay process of conical singularities
to the plane. Thus, Ricci flows provide a non-linear generalization 
of the heat equation in two dimensions with the same dissipative properties.     
Various applications to dynamical problems of string theory are also briefly
discussed. Finally, we outline generalizations to higher dimensional target 
spaces that exhibit sufficient number of Killing symmetries.  
\vfill
\end{titlepage}
\eject

\def\baselinestretch{1.2}
\baselineskip 16 pt
\noindent

The renormalization group properties of two-dimensional sigma models were extensively
studied in the past, and it was found that their beta function $\beta (g^2)$  
is not always zero (see, for instance, \cite{sasha}). In general, the coupling constant
runs by changing the length scale of the world-sheet and the theory breaks conformal
invariance in the quantum regime. This general framework has two applications in 
string theory. First, it is interesting to find fixed points of the renormalization 
group flow, as they provide conformal field theory blocks for building string vacua. 
Second, it is also interesting to study non-conformal backgrounds by solving the
beta function equations and try to develop an off-shell formulation for addressing
the problem of vacuum selection in string theory. Our work is mostly concerned 
with the second problem, as we report on a new algebraic method for the 
integration of the renormalization group equations that appear to lowest order in 
perturbation theory. Further details can be found in a recent publication, 
\cite{bakas}.      
     
We set us the notation by considering two-dimensional sigma models with classical  
action
\be
S = {1 \over 4\pi \alpha^{\prime}} \int d^2w \sqrt{h} h^{ij} (\partial_i X^{\mu}) 
(\partial_j X^{\nu}) G_{\mu \nu} ~,  
\ee
where $\{X^{\mu}; \mu = 1, 2, \cdots , n\}$ are coordinates in target space with 
metric $G_{\mu \nu} (X)$. Let $\Lambda^{-1}$ be the renormalization scale parameter
on the world-sheet and $R_{\mu \nu}$ the Ricci curvature tensor of the target space
metric. Then, the renormalization group equations are
\be
\Lambda^{-1} {\partial \over \partial \Lambda^{-1}} G_{\mu \nu} = 
- \beta (G_{\mu \nu}) = -R_{\mu \nu} + \cdots 
\ee
in appropriate units, $2\pi \alpha^{\prime} = 1$, and dots denote higher order 
curvature terms that arise at two loops or higher in perturbation theory, \cite{dan}. 
We will only consider the lowest order terms, which provide a good approximation    
to the metric beta functions 
when the curvature is rather weak, and define the logarithmic length scale 
$t = {\rm log} \Lambda^{-1}$ as the {\em renormalization group time}; in this 
parameterization, the ultra-violet limit corresponds to $t \rightarrow -\infty$ and
the infra-red limit to $t \rightarrow +\infty$. Thus, we consider the equations
\be
{\partial \over \partial t} G_{\mu \nu} = -R_{\mu \nu} ~, \label{flows}  
\ee
which define a dynamical system in superspace that consists of all possible metrics.

These equations describe changes of the target space geometry that are induced by 
changes of the world-sheet logarithmic scale, but they are very difficult to solve
in all generality. The same equations also arose independently in the mathematics 
literature, where they became known as {\rm Ricci flows}, and they have turned  
into a major tool for addressing a variety of open problems in geometry in diverse 
dimensions; for a recent account of the main mathematical results see, for 
instance, \cite{ricci}, and references therein. We will be able to solve these
equations for two-dimensional target spaces by casting them into zero 
curvature form and then integrate them in closed form. As byproduct, we 
reveal the role of a novel infinite dimensional Lie algebra in this dynamical 
problem, which arises in the zero curvature formulation and it could be of 
more general value. Some special trajectories will also be briefly reviewed in
order to confront the algebraic construction of the general solution with the 
geometric deformations that are involved. We also note for completeness that 
the renormalization group equations \eqn{flows} admit a more general 
description in the form
\be
{\partial \over \partial t} G_{\mu \nu} = -R_{\mu \nu} 
+ \nabla_{\mu} \xi_{\nu} + \nabla_{\nu} \xi_{\mu} ~, \label{flows2}  
\ee
which incorporates the effect of all possible field redefinitions 
$\delta X_{\mu} = -\xi_{\mu}$ associated with changes of coordinates along 
the flow. Some solutions assume a simpler form when they are written in 
different frames.     

Let us now restrict attention to two-dimensional target spaces in the conformally 
flat frame 
\be
ds_{\rm t}^2 = 2 e^{\Phi(z_+, z_-; t)} dz_+ dz_- = {1 \over 2} e^{\Phi(X,Y; t)} 
(dX^2 + dY^2)
\ee
using Cartesian coordinates $X$, $Y$ or the complex conjugate variables 
$2z_{\pm} = Y \pm i X$. Then, since the non-vanishing component of the Ricci 
curvature tensor equal to
\be
R_{+-} = -\partial_+ \left(e^{-\Phi} \partial_- e^{\Phi} \right) = - \nabla^2 
\Phi ~,  
\ee
we find that the renormalization group flows \eqn{flows} reduce in this frame to the 
following non-linear differential equation for $\Phi(X, Y; t)$,  
\be
{\partial \over \partial t} e^{\Phi(X, Y; t)} = \nabla^2 \Phi(X, Y; t) ~. 
\label{main} 
\ee
  
This equation provides a non-linear generalization of the heat equation on the 
two-dimensional plane, 
\be
{\partial \over \partial t} \Theta(X, Y; t) = \nabla^2 \Theta (X,Y; t) ~,  
\ee
which approximates \eqn{main} when $\Phi(X, Y; t) \simeq \Theta (X,Y; t)$ becomes 
very small. Such linearization typically arises in the infra-red region 
of the renormalization group flow, $t \rightarrow +\infty$, where the target
space metric becomes approximately flat and the weak field limit  
\be
G_{\mu \nu} \simeq \delta_{\mu \nu} + h_{\mu \nu} ~~~~~~~ {\rm with} ~~~ 
h_{\mu \nu} <<1 ~~~ {\rm and} ~~~ \partial h_{\mu \nu} <<1 \label{weak}  
\ee
comes into play. The heat equation exhibits dissipative behavior in time, $t$,
which can be seen by considering the fundamental solution
\be
\Theta(X, Y; t) = {1 \over 4\pi t} e^{-(X^2 + Y^2)/4t} 
\ee
that describes a Gaussian pulse with height $1/t$ and width $\sqrt{t}$. It starts
as singular function at the initial time $t=0$, where the Gaussian  
becomes delta function, and it dissolves after infinitely long time by spreading
the initial singularity all over the plane. However, since the weak field 
approximation \eqn{weak} is not valid for finite values of $t$, 
but only asymptotically as $t \rightarrow +\infty$,
one may wonder whether the characteristic property of the heat equation to
dissipate singularities is also a property of the non-linear equation 
\eqn{main}. 
As we will see later, the decay process of a conical singularity, which 
describes the transition of a flat cone $C/Z_n$ to the plane, provides 
an analogue of the fundamental solution of the heat equation 
to the non-linear system \eqn{main}.   

The main equation \eqn{main} can be conveniently written in the form 
\be
\nabla^2 \Phi(X, Y; t) = \int dt^{\prime} K(t, t^{\prime}) 
e^{\Phi(X, Y; t^{\prime})} ~, 
\ee
where $K(t, t^{\prime})$ is the kernel
\be
K(t, t^{\prime}) = {\partial \over \partial t} \delta (t, t^{\prime}) ~,
\label{cartan}  
\ee
which is anti-symmetric with respect to its arguments $t$ and $t^{\prime}$. 
This formulation proves advantageous for casting the Ricci flow into 
Toda form for appropriate choice of the Lie algebra. Recall at this point 
that Toda equations describe integrable interactions for a collection of
two-dimensional fields $\{\phi_i (X, Y)\}$, which are labeled by the simple 
roots of a given Lie algebra and they are coupled via the Cartan matrix 
$K_{ij}$ as follows,
\be
\nabla^2 \phi_i(X,Y) = \sum_j K_{ij} e^{\phi_j(X,Y)} ~. 
\ee
The indices $i$, $j$ are typically discrete, but there are also generalizations 
to continuous variables for some infinite dimensional Lie algebras, which are
obtained by replacing the collection of two-dimensional Toda fields 
$\{\phi_i\}$ with a ``master" field $\Phi(X, Y; t)$, the Cartan matrix $K_{ij}$ 
with a Cartan kernel $K(t, t^{\prime})$, and the summation over $j$ with an 
integral over $t^{\prime}$. In this case, $t$ is interpreted as Dynkin parameter
on the root system of a continual Lie algebra whose basic commutation relations
are defined by the corresponding Cartan kernel, which depends on continuous 
variables rather than discrete labels. The Ricci flows in two dimensions 
fit precisely this algebraic framework provided that $K(t, t^{\prime})$ is 
given by \eqn{cartan} with $t$ being the logarithm of the world-sheet length scale 
of sigma models. Its anti-symmetry is inherited from the first order derivatives
that describe the evolution in $t$.    

The notion of continual Lie algebras arises naturally by considering a system of
Cartan-Weyl generators $\{H(t), X^{\pm}(t)\}$ that depend on a continuous 
variable $t$, so that the commutation relations of the local part of the algebra
assume the general form, \cite{misha}  
\ba
& & [X^+(t) , X^-(t^{\prime})] = \delta(t-t^{\prime}) H(t^{\prime}) ~, ~~~~~~ 
[H(t), H(t^{\prime})] = 0 ~, \nonumber\\
& & [H(t), X^{\pm}(t^{\prime})] = \pm K(t, t^{\prime}) X^{\pm} (t^{\prime}) ~. 
\ea
Equivalently, using the smeared form of the generators 
$\{H(\varphi), X^{\pm}(\varphi)\}$ with respect to arbitrary function $\varphi(t)$,
where
\be
A(\varphi) = \int dt \varphi(t) A(t) ~,  
\ee
the commutation relations above assume the form
\ba
& & [X^+(\varphi) , X^-(\psi)] = H(\varphi \psi) ~, ~~~~~~ 
[H(\varphi), H(\psi)] = 0 ~, \nonumber\\
& & [H(\varphi), X^{\pm}(\psi)] = \pm X^{\pm} (K(\varphi, \psi)) ~. 
\ea
Here, we restrict attention to bilinear maps 
$K(\varphi, \psi) = (K\varphi) \cdot \psi$, which are described by a much simpler 
linear map $K$, so that there is always consistency with the Jacobi identities.
The algebra that results in this case is denoted by ${\cal G}(K, 1)$ and it is 
$Z$-graded in the sense that
\be
{\cal G}(K, 1) = \oplus_{n \in Z} {\cal G}_n ~. 
\ee
The local part corresponds to ${\cal G}_{-1} \oplus {\cal G}_0 \oplus {\cal G}{+1}$, 
whereas all other elements are obtained, as usual, by taking successive commutators
so that ${\cal G}_n = [{\cal G}_{n-1}, {\cal G}_{+1}]$ for $n>0$ and 
${\cal G}_n = [{\cal G}_{n+1}, {\cal G}_{-1}]$ for $n<0$.    

The Ricci flows admit a zero curvature formulation as Toda theory for the special  
continual Lie algebra with $K = d/dt$, \cite{bakas}. It can be easily seen by considering
\be
[\partial_+ + A_+(z_+, z_-) , ~ \partial_- + A_-(z_+, z_-)] = 0 ~, 
\ee
where the gauge connections $A_{\pm}$ take values in the local part of the 
algebra ${\cal G}_0 \oplus {\cal G}_{\pm 1}$, respectively, with 
\be
A_{\pm}(z_+, z_-) = H(u_{\pm}) + \lambda X^{\pm} (f_{\pm}) 
\ee
that depend on some arbitrary functions $u_{\pm}$ and $f_{\pm}$ of $z_+$, 
$z_-$ and $t$. Then, using the basic commutation relations of the algebra
${\cal G}(d/dt; 1)$ we obtain the following system of equations
\ba
& & {\partial u_+ \over \partial t} = -\partial_+ ({\rm log}f_-) ~, ~~~~~ 
{\partial u_- \over \partial t} = \partial_- ({\rm log} f_+) ~, 
\nonumber\\
& & \partial_+ u_- - \partial_- u_+ + \lambda^2 f_+ f_- = 0 ~, 
\ea
which can be simplified by eliminating $u_{\pm}$. Then, the zero curvature 
condition yields
\be
\partial_+ \partial_- \Phi(z_+, z_-; t) = -\lambda^2 {\partial \over \partial t}
e^{\Phi(z_+, z_-; t)} ~, ~~~~~~ {\rm with} ~~~ \Phi = {\rm log}(f_+ f_-) ~, 
\ee
which is the Ricci flow equation \eqn{main} in conformally flat coordinates $z_+$, 
$z_-$ with $\lambda^2 = -1$. Equivalently, using the gauge invariance of the
zero curvature condition, we may choose $u_- =0$, $f_+ =1$ from the very beginning,   
and set $f_- = {\rm exp} \Phi$, $u_+ = \Psi$, where $\partial_+ \Phi = -  
\partial_t \Psi$. Thus, the renormalization group equations are integrable in 
target space, while $t$ enters into the definition of the infinite dimensional
Lie algebra that describes them as continual Toda system.      

According to the general theory of Toda field equations, the general 
solution can be obtained in closed form, using B\"acklund transformations, and
it is parametrized by a system of arbitrary free fields. The group theoretical
method that allows for their integration is based on the existence of a 
highest weight state $|t>$, so that  
\be
X^+(t^{\prime})|t> = 0 ~, <t|X^-(t^{\prime}) = 0 ~, 
H(t^{\prime}) |t> = \delta(t-t^{\prime}) |t> 
\ee
with $<t|t> = 1$. This is formally defined for continual Lie algebras, in 
analogy with the representations of finite dimensional simple Lie algebras.
Then, for the case at hand, the general solution assumes the form, 
\cite{bakas}  
\be
\Phi(z_+, z_-; t) = \Phi_0(z_+, z_-; t) + \partial_t \left({\rm log} 
<t| M_+^{-1}(z_+; t) M_-(z_-; t)|t> \right) , 
\ee
where $M_{\pm}$ are path-ordered exponentials  
\be
M_{\pm} (z_{\pm}; t) = {\cal P} {\rm exp} \left(i \int^{z_{\pm}} 
dz_{\pm}^{\prime} \int^t dt^{\prime} e^{f^{\pm}(z_{\pm}^{\prime}; t^{\prime})} 
X^{\pm} (t^{\prime}) \right) 
\ee
and $\Phi_0 (z_+, z_-; t) = f^+ (z_+; t) + f^-(z_-; t)$ is a one-parameter family of
free fields in two dimensions, i.e., $\partial_+ \partial_- \Phi_0 (z_+, z_-; t) = 0$, 
which depend on the continuous index $t$.  

The solution can be written in formal power series around the free field configuration 
by expanding the path-ordered exponentials, so that  
\ba
& & <t|M_+^{-1} M_- |t> = 1 + \sum_{m=1}^{\infty} \int^{z_+} dz_1^+ \cdots 
\int^{z_{m-1}^+} dz_m^+ \int^{z_-} dz_1^- \cdots \int^{z_{m-1}^-} dz_m^- \times 
\nonumber\\
& & ~~~~~~~~ \times \int \prod_{i=1}^m dt_i \int \prod_{i=1}^m dt_i^{\prime} 
{\rm exp} f^+(z_i^+; t_i) {\rm exp}f^-(z_i^-; t_i^{\prime}) 
D_t^{\{t_1, \cdots , t_m; t_1^{\prime}, \cdots , t_m^{\prime}\}} ~. 
\ea 
Here, the matrix elements  
\be
D_t^{\{t_1, t_2, \cdots , t_m; t_1^{\prime}, t_2^{\prime}, \cdots , t_m^{\prime}\}} 
= <t| X^+(t_1) X^+(t_2) \cdots X^+(t_m) X^-(t_m^{\prime}) \cdots X^-(t_2^{\prime}) 
X^-(t_1^{\prime}) |t>  
\ee
can be evaluated one by one, using the basic commutation relations of the algebra 
${\cal G}(d/dt; 1)$, and their integration against the free field $f^{\pm}$ 
yields all terms in the free field expansion of the Toda field configuration 
$\Phi(z_+, z_-; t)$. The details are rather cumbersome, since the generic term of
the expansion can not be written in closed form, but for special configurations
that correspond to free fields 
\be
\Phi_0(z_+, z_-; t) = c \cdot (z_+ + z_-) + d(t) \equiv cY + d(t) 
\ee
the result is more manageable. Such configurations are used to parametrize solutions
of the Toda field equation with an isometry, i.e., independent of $X$, and they give
rise to the following perturbative expansion, \cite{bakas} 
\be
\Phi = \Phi_0 + {1 \over c^2} \partial_t e^{\Phi_0} + {1 \over 4c^4} \partial_t 
\left(e^{\Phi_0} \partial_t e^{\Phi_0} \right) + {1 \over 36c^6} \partial_t 
\left(3e^{\Phi_0} \left(\partial_t e^{\Phi_0} \right)^2 + e^{2\Phi_0} 
\partial_t^2 e^{\Phi_0} \right) + \cdots \label{free}  
\ee
which is valid around ${\rm exp} \Phi \simeq 0$, i.e., for 
$\Phi_0 \rightarrow -\infty$.    

The infinite dimensional algebra ${\cal G}(d/dt; 1)$ has exponential growth, which
is seen by attempting to construct the independent elements that parametrize the
subspaces ${\cal G}_n$ beyond the local part $n=0, \pm 1$. It can be shown by 
induction that if ${\cal G}_{\pm n}$ is spanned by $d_n$ independent elements
$X_{\pm n}^{(1)}, \cdots , X_{\pm n}^{(d_n)}$, the subspace 
${\cal G}_{\pm (n+1)} = [{\cal G}_{\pm 1} , {\cal G}_{\pm n}]$ will be spanned 
by the following $2d_n$ independent elements, \cite{misha} 
\ba
X_{\pm (n+1)}^{(s)} (\varphi) & = & \alpha_n^{(s)} [X_{\pm 1}(1), 
X_{\pm n}^{(s)} (\varphi)] - [X_{\pm 1} (\varphi), X_{\pm n}^{(s)} (1)] ~; 
~~~~ 1 \leq s \leq d_n ~, \\
X_{\pm (n+1)}^{(s)} (\varphi) & = & \beta_n^{(s)} [X_{\pm 1}(1), 
X_{\pm n}^{(s)} (\varphi)] + [X_{\pm 1} (\varphi), X_{\pm n}^{(s)} (1)] ~; 
~~~~ d_n + 1 \leq s \leq d_n 
\ea
for all $n \geq 2$ and for appropriately chosen constants $\alpha_n^{(s)}$ and 
$\beta_n^{(s)}$. This formula does not apply for $n=1$, since the two 
different series of elements are linearly dependent and 
each ${\cal G}_{\pm 2}$ has only
one generator. Thus, the dimension of the subspaces 
${\cal G}_{\pm n}$ 
(relative to the dimension of ${\cal G}_0$, which is taken to be 1) turns out to
be 
\be
d_0 = d_{\pm 1} =1 ~, ~~~~~ d_n = 2^{n - 2} ~~~~ {\rm for} ~~~ 
n \geq 2 
\ee
and ${\cal G}(d/dt; 1)$ exhibits exponential growth. This makes it rather exotic and
also difficult to study in great detail. The complete system of commutation relations
is not known in closed form, but fortunately we only need the local part 
${\cal G}_{-1} \oplus {\cal G}_0 \oplus {\cal G}_{+1}$ to write down the Ricci flows
in Toda form. It should be noted, however, that generalized systems 
of Toda field equations can also be written down in zero curvature form 
$[\partial_+ + A_+, ~ \partial_- + A_-] = 0$ for appropriately chosen gauge connections
$A_{\pm} \in {\cal G}_0 \oplus {\cal G}_{\pm 1} \oplus \cdots \oplus {\cal G}_{\pm N}$ 
for all $N$. It is natural to expect that such equations will be associated to the 
beta functions of higher spin fields beyond the renormalization group equation of the 
target space metric.       

Returning back to the geometry of Ricci flows, we present some special solutions that
have received a lot of attention in recent years. The first describes axially symmetric 
deformations of the round sphere, which look like a sausage that becomes infinitely  
long in the ultra-violet region, \cite{sausa}; the flow terminates   
at a finite scale, say $t=0$, where the configuration collapses to a point and 
the lowest order approximation to the beta function equations is not valid anymore. 
The second describes the decay process of a conical singularity by considering 
an orbifold configuration $C/Z_n$ at some initial time, say $t=0$, which then  
flows to flat space in the infra-red region, \cite{cone1, cone2}. 
Both examples are tractable and they can be 
derived by much simpler integration methods, because they 
correspond to mini-superspace approximation of the Ricci flow equations 
imposed by axial symmetry. 
Other special solutions are also known in the 
literature, \cite{bakas}, and they all fit into the general algebraic scheme that
was presented above.

(i) {\em Sausage model}: In this case, the conformal factor of the metric assumes
the simple form
\be
e^{\Phi (Y; t)} = {2 \over a(t) + b(t) {\rm cosh} 2Y} ~, 
\ee
where $0 \leq X \leq 2\pi$ and $-\infty < Y < +\infty$. Then, the flows \eqn{flows} 
reduce to the first order system of differential equations
\be
a^{\prime}(t) = 2 b^2(t) ~, ~~~~~~ b^{\prime} (t) = 2a(t)b(t) ~, 
\ee
which are easily solved in the physical region $a(t) \geq b(t) \geq 0$ as follows, 
\be
a(t) = \gamma {\rm coth}(-2\gamma t) ~, ~~~~~ b(t) = 
{\gamma \over {\rm sinh}(-2\gamma t)}  ~, 
\ee
with $\gamma$ being an arbitrary non-negative constant and $t$ running from 
$-\infty$ (ultra-violet) to $0$ (big crunch). This special solution admits 
a free field representation, in the context of equation \eqn{free}, provided that
\be
\Phi_0 (Y; t) = -2Y + {\rm log} \left({4 \over \gamma} {\rm sinh}(-2\gamma t) 
\right) ~. 
\ee
  
It can also be written in proper coordinates, where it becomes simpler to
visualize as sausage, 
\be
ds_{\rm t}^2 = {k \over \gamma} \left( d\tilde{Y}^2 + {\rm sn}^2 
\left(\tilde{Y} + K(k) ; k \right) dX^2 \right) , 
\ee
by introducing change of coordinates $\tilde{Y} = F(\psi; k)$ given by the 
incomplete elliptic integral of the first kind with free parameter 
${\rm sin} \psi = {\rm tanh} Y$ and modulus $k = {\rm tanh}(-\gamma t)$.  
In this frame, however, one has to introduce a compensating vector field
$\xi_{\mu} = \partial_{\mu} \tilde{\Phi}$ with 
\be
\tilde{\Phi} (\tilde{Y}) = {\rm log} \Theta (\tilde{Y} + K(k))  + 
{1 \over 2} \left({E(k) \over K(k)} - {1 \over 2} {k^{\prime}}^2 \left(1 + 
{1 \over \gamma} \right) \right) {\tilde{Y}}^2   
\ee
following equation \eqn{flows2}. All 
formulae are written here in terms of the standard Jacobi elliptic and theta 
functions. As for the solution, it can also be viewed as bound state of 
two Euclidean two-dimensional 
black holes which are glued together in the asymptotic region and they evolve by 
``eating" each other until they reach a singularity at $t=0$. 
  
(ii) {\em Decay of cone}: This process is better described in a frame where 
the $t$ dependence factorizes linearly, as 
\be
ds_{\rm t}^2 = t \left(f^2(r) dr^2 + r^2 d\phi^2 \right) , ~~~~~ 
\xi_r = {1 \over 2}rf(r) ~, ~~~~~ \xi_{\phi} = 0 ~,  
\ee
where $0 \leq \phi \leq 2\pi /n$ 
and the vector field $\xi_{\mu}$ accounts for the formulation of the 
solution in a  non-conformally flat system. Then, the renormalization group 
equations \eqn{flows2} result into a simple differential equation for $f(r)$ with 
solution
\be
\left({1 \over f(r)} - 1 \right) {\rm exp} \left({1 \over f(r)} -1 \right) 
= (n-1) {\rm exp} \left(n-1 - {r^2 \over 2} \right)  
\ee
that interpolates smoothly between $f=1/n$ at $r=0$ and $f=1$ at $r = \infty$. 
Equivalently, introducing the $t$-dependent change of coordinates $\tilde{r} 
= r \sqrt{t}$, we find that when $\tilde{r}$ is held fixed the limit 
$t \rightarrow 0$ corresponds to $f \rightarrow 1$ and $t \rightarrow +\infty$ 
corresponds to $f \rightarrow 1/n$. Thus, the geometry describes the transition
from a cone $C/Z_n$, which has opening angle $2\pi / n$ at the initial time, to
the plane which is only reached in the infra-red limit. The solution looks more
complicated when formulated in a conformally flat frame, but it can be worked 
out in detail and compared with the general solution of the Toda field 
equation for appropriate choice of free field configuration, \cite{bakas}. 

The decay process of a conical singularity demonstrates by example the dissipative
nature of the Ricci flows, which are capable to spread an initial
curvature singularity all over space after infinitely long time; 
as such, it can be viewed as non-linear generalization of 
the fundamental Gaussian solution of the heat equation. This process, which is 
described in the gravitational regime, has several applications in closed string 
theory while studying the general problem of tachyon condensation. For instance,
we may consider ten-dimensional string vacua of the form
\be
{\cal C} = {\rm CFT}_2 + R^{7, 1} ~, 
\ee
where the conformal block ${\rm CFT}_2$ corresponds to the orbifold field 
theory $C/Z_n$. This vacuum is unstable because there are tachyonic states in the
twisted sector of the orbifold, which are localized at the tip of the cone, and
they induce transitions to more stable vacua by reducing the order of the 
singularity; the singularity is completely resolved by the decay of the cone to the 
plane, as given by the above geometric transition in renormalization group time.    
More generally, tachyon condensation is quite important for addressing the 
problem of vacuum selection in string theory by dynamical methods. Ricci flows
offer a good starting point in this direction. 

The deformations of two-dimensional geometries, which are induced by Ricci flows,  
could be interpreted as time dependent gravitational solutions in $2 + 1$ 
dimensions, which are on-shell.  
It should be noted, however, that the correspondence with dynamical processes in
real time has not been made precise in all generality to this day, although 
special examples that exhibit the same qualitative behavior with the 
renormalization group flows are known in the literature, \cite{harvey}. 
It is expected that 
a similar dynamical interpretation will be possible in any number of 
dimensions, and could be used either way. 
This formulation is quite important for addressing a variety of physically
interesting dynamical problems, such as gravitational collapse, in terms of a 
more fundamental standpoint provided by world-sheet methods.  
There are critical phenomena associated to gravitational collapse 
(for a review see, for instance, \cite{golm}), which could be better understood
in terms of renormalization group flows of $d$-dimensional geometries rather than      
using dynamical processes in $(d+1)$-dimensional space-time. In this context, it
will also be interesting to revisit Zamolodchikov's $c$-theorem, \cite{zamo}, and 
its effective field theory description as entropy of the Ricci flows, following
the recent work of Perelman, \cite{grisha}.        
 
Another interesting problem is the generalization of our algebraic framework to  
higher dimensional target spaces. The formulation of Ricci flows as zero curvature 
condition is rather special to two dimensions, but it can be extended to 
higher dimensional geometries with sufficient number of Killing isometries, so that
the metric is effectively described by functions of two coordinates. 
For example, in three 
dimensions, we may consider the action of Ricci flows taking place in the restricted 
class of metrics with Killing coordinate $w$, 
\be
ds_{\rm t}^2 = V(z_+, z_-; t) dw^2 + 2 e^{\Phi (z_+, z_-; t)} dz_+ dz_- ~, 
\ee
and try to describe the evolution equations for the functions $V$ and $\Phi$ in 
a similar way. In this context, the decay of conical geometries might admit 
a nice interpretation in terms of matter sources coupled to gravity. The 
systematic extension of our work to all these cases remains open as there are 
several two-dimensional fields, and not just $\Phi$, which come into play. 
We hope to be able to report elsewhere whether they all fit in a generalized 
Toda system, as for two-dimensional geometries.    
  
In summary, the Ricci flows provide a meeting 
point for many recent developments in physics and mathematics, which could be used  
further for the benefit of both subjects. Our main contribution in this area is
the algebraic formulation of two-dimensional Ricci flows as Toda system, which in
turn points to a new class of infinite dimensional algebras that are relevant for
addressing dynamical problems.

\newpage
\centerline{\bf Acknowledgments}
\noindent
This work was supported in part by the European Research and Training Networks
``Superstring Theory" (HPRN-CT-2000-00122) and ``Quantum Structure of 
Space-time" (HPRN-CT-2000-00131), as well as the Greek State Foundation Award
``Quantum Fields and Strings" (IKYDA-2001-22) and NATO Collaborative Linkage
Grant ``Algebraic and Geometric Aspects of Conformal Field Theories and 
Superstrings" (PST.CLG.978785). I also thank the organizers of the meeting 
for their kind invitation to present the results of my work in a pleasant 
and stimulating atmosphere.


\begin{thebibliography}{3}
\bibitem{grisha}
G. Perelman, ``The entropy formula for the Ricci flow and its geometric 
applications", math.DG/0211159.
\end{thebibliography}
\end{document}